\documentclass[preprint,12pt]{elsarticle}

 \usepackage{graphicx}
\usepackage{amssymb}
\usepackage{amsmath}

\journal{Physics Letters B}

\begin{document}

\begin{frontmatter}



\title{Small $x$ resummation and the Odderon}


\author{Anna M. Sta\'sto}

\address{Physics Department, Penn State University, University Park, 16802 PA, U.S.A.\fnref{ff} }
\address{RIKEN center, Brookhaven National Laboratory, Upton, 11973 NY, U.S.A.}
\address{Institute of Nuclear Physics, Polish Academy of Science, ul. Radzikowskiego 152, Krak\'ow, Poland}
\begin{abstract}
We present a general argument which suggests that the Bartels-Lipatov-Vacca Odderon intercept should be equal to one to all orders in the perturbation theory.  The argument is based on the validity of the  so called omega-expansion in the high energy limit. It  can be further supported by the analogous pattern observed in the case of the anomalous dimensions which is a consequence of the momentum sum rule. In addition, we conjecture that the BFKL kernel should satisfy the transverse momentum sum rule. Finally, it is shown that the higher order kinematical effects do not change the BLV Odderon intercept.
\end{abstract}


\end{frontmatter}


\section{Introduction}
\label{sec:intro}

The Odderon is the $C$ - odd partner of the Pomeron, and in QCD it can be represented as a 3-gluon color singlet state. The small $x$ evolution for the Odderon was derived long time ago \cite{Bartels:1978fc,Bartels:1980pe,Kwiecinski:1980wb}. Two solutions for the intercept of the Odderon are known. The first one, the Janik-Wosiek solution \cite{Janik:1998xj}, was obtained from the requirement of the holomorphic symmetry, which has intercept slightly less than unity. The second one, which we will call the BLV Odderon (Bartels-Lipatov-Vacca) \cite{Bartels:1999yt} has intercept exactly equal to unity (this  was also confirmed later in \cite{Kovchegov:2003dm} and \cite{Hatta:2005as}). This solution has a special feature, namely that two out of three gluons are in the same position in the transverse plane. Since it is $C$-odd it is found by selecting odd conformal spins (denoted further  by $n$)  from the  spectrum of the leading order BFKL Pomeron eigenvalue \cite{BFKL}. Thus the Odderon intercept is dominated by $n=1$ conformal spin for which it is exactly equal to unity.

An interesting question arises \cite{Jochen}, whether the BLV Odderon intercept still has intercept equal to unity
when the higher order terms are taken into account. The BFKL Pomeron is known to NLLx order \cite{Fadin:1998py,Kotikov:2000pm,Camici:1997ij}, and the corrections are numerically large.

For the Odderon the higher order corrections do not have to be large. Roughly speaking one can argue that because this antisymmetric solution selects the odd conformal spins, the transverse momentum integrals in the kernel should vanish when  integrated with the eigenfunctions for which $n=1$ and $\gamma=1/2$, where $\gamma$ is the conjugated variable to the transverse momentum.  Little more care is needed when considering the running coupling, but a recent analysis \cite{Braun:2007kz}, demonstrated that the BLV Odderon intercept is not affected. 

In this letter, we construct a different argument, which is based on the assumption of the so-called $\omega$ expansion \cite{Ciafaloni:1998iv,Ciafaloni:1999yw}. Here $\omega$ is the variable Mellin conjugated to the energy $s$. This expansion was used  to construct the resummation scheme for the case of the BFKL Pomeron. The basic statement is that in the high energy limit  $\omega$ is more natural expansion parameter  than $\alpha_s$.  The lowest order terms in $\omega$ and $\alpha_s$ expansions are identical. Then,   in the case when the lowest order is vanishing, the requirement of the existence of such expansion implies that all the higher orders in  $\alpha_s$ expansion are vanishing too. 

This argument can be further supported by the analysis of the anomalous dimensions. There, a dual $\gamma$ expansion can be constructed, which is suitable for the resummation in the collinear regime\footnote{The duality between anomalous dimensions and eigenvalues was the basis of another resummation approach \cite{Altarelli:1999vw}.}. Since the momentum sum rule forces the anomalous dimensions to vanish to all orders when $\omega \rightarrow 1$,
the $\gamma$ expansion holds.

We also check that the NLL BFKL kernel vanishes at $n=1$ and $\gamma=1/2$ which
is consistent with the $\omega$ expansion.  We suggest that the vanishing of the BFKL kernel at this particular point can be associated with the momentum sum rule for the transverse components of the momenta. 

In the next section we formulate the consistency argument based on the $\omega$ expansion. We  conjecture that the BFKL should satisfy a separate momentum sum rule for the transverse components of the momenta.
In Sec.~\ref{sec:nllx} we check that the  NLLx BFKL intercept calculated from \cite{Fadin:1998py} vanishes for
 $n=1$ and is thus consistent with the conjectured transverse momentum sum rule the  whereas the result \cite{Balitsky:2008zz} apparently does not vanish.
Finally in Sec.~\ref{sec:oddkc} we investigate the Odderon eigenvalue with the pole shifts due to the kinematical constraints and show that the intercept is still equal to unity, however the diffusion is  significantly reduced.

\section{Consistency argument of  the $\omega$ expansion}
\label{sec:omegaexp}

 We consider a  
 BFKL equation\cite{BFKL} in the $\omega$ representation
 \begin{equation}
\omega \, G_{\omega}(k_T,k_{T}^0) =  \delta^{(2)}(k_T,k_T^0) + \int \frac{d^2 k'_T}{2\pi} K(k_T,k_T') \, G_{\omega}(k_T',k_T^0) \; ,
\label{eq:bfkl1}
\end{equation}
 where the integral is over the transverse momentum $k_T$. Here, $ G_{\omega}$ is the gluon  Green's function, with $\omega$ being the variable Mellin conjugated to logarithm of the energy. The solution to the above equation is found by performing Mellin transform in $k_T$ where one finds the  condition
  \begin{equation}
 \omega(n,\gamma,\alpha_s) =  \bar{\alpha}_s \chi(\gamma,n) \; ,
  \label{eq:eigcond}
 \end{equation}
 where $\chi(\gamma,n)$ is the BFKL kernel eigenvalue, $n$ is conformal spin, and $\gamma$ is the Mellin variable conjugated to $\ln k_T/k_T^0$. Strong coupling is redefined to be $\bar{\alpha}_s=\alpha_s N_c/\pi$.

The kernel eigenvalue can be expanded in the powers of the strong coupling
\begin{equation}
\chi(n,\gamma) =  \chi_0(n,\gamma) +\sum_{k \ge 1} \bar{\alpha}_s^{k} \, \chi_{k}(n,\gamma) \; ,
\label{eq:alphasexpand}
\end{equation}
and using perturbation theory one can find corresponding functions $\chi_{k}(n,\gamma)$ which are independent of $\omega$. Up to now they are known for  $k=0,1$. 

The concept of the  $\omega$ expansion \cite{Ciafaloni:1998iv,Ciafaloni:1999yw}  can be argued as follows. The Regge limit is defined as the asymptotic limit in which $s \gg |t| \gg \Lambda^2$, that is  the center of mass energy is much larger (essentially infinite) than the other scales which characterize the scattering process. 
 The strong coupling  though is  a parameter which is not necessarily small in this limit. In that case the convergence of a perturbative series in $\alpha_s$ as given by (\ref{eq:alphasexpand}) might not be very fast, which is indeed the case given the size of the NLLx corrections. This leads to the hypothesis that the parameter which  better characterizes the expansion in the high energy limit is $\omega$, which should be small  in this case.

Therefore one makes the ansatz that the  eigenvalue of the kernel  in Eq.~\ref{eq:bfkl1} posses the following representation\begin{equation}
\tilde{\chi}(n,\gamma,\omega) = \tilde{\chi}_0(n,\gamma) + \sum_{j\ge 1} \omega^j \tilde{\chi}_j(n,\gamma,\omega=0) \; ,
\label{eq:kerneltilde} 
\end{equation}
with 
\begin{equation}
\omega = \bar{\alpha}_s \tilde{\chi}(n,\gamma,\omega) \;.
\label{eq:implicit}
\end{equation}

Obviously the equation above should give the same value  for the resulting intercept. Therefore the two functions $\chi(n,\gamma)$ and $\tilde{\chi}(n,\gamma,\omega)$ are both  equal to each other and to the intercept.
The difference is that now since the kernel eigenvalue itself depends on $\omega$ one needs to solve the complicated nonlinear equation in the form (\ref{eq:implicit}).

Now, given the expansion (\ref{eq:alphasexpand}) one can get the second one (\ref{eq:kerneltilde}) which leads to relation
\begin{multline}
\tilde{\chi}(n,\gamma,\omega) \, = \, {\chi}_0(n,\gamma) +\omega \frac{\chi_1(n,\gamma)}{\chi_0(n,\gamma)} \, +\\
+\,\omega^2 \frac{1}{\chi_0(n,\gamma)}\left( \frac{\chi_2(n,\gamma)}{\chi_0(n,\gamma)}-\left(\frac{\chi_1(n,\gamma)}{\chi_0(n,\gamma)}\right)^2\right)\,+\,{\cal O}(\omega^3) \; ,
\label{eq:omegaexp}
\end{multline}
which is the $\omega$ expansion \cite{Ciafaloni:1998iv,Ciafaloni:1999yw}.
Strictly speaking the above formula is valid for the case of the fixed coupling.
It can be generalized to include the effects of the running coupling in which case the series
is modified by terms  proportional to the $\beta$ function and include the differential operator $\partial_{\gamma}$.

Additionally we have the lowest order condition, because two terms in the leading order in both expansions have to coincide with each other
\begin{equation}
\chi_0(n,\gamma) \, \equiv \, \tilde{\chi}_0(n,\gamma) \;.
\label{eq:llx}
\end{equation}
When going to higher  orders in each fixed order in $\alpha_s$ we have corresponding string of terms
with different orders in $\omega$, see \cite{Ciafaloni:1998iv,Ciafaloni:1999yw}.

As it is clear from (\ref{eq:omegaexp}) this expansion could become  invalid when the eigenvalue $\chi_0(n,\gamma)\rightarrow 0$.  This is precisely the case for the BLV Odderon \cite{Bartels:1999yt} since at the lowest order
$$
\omega_0=\bar{\alpha}_s \chi_0(n=1,\gamma=1/2)=0 \; .
$$
  Therefore, at first sight,  this would indicate a complete failure of the $\omega$ expansion argument. There is however a distinct possibility for the $\omega$ expansion to be reliable even in this case, provided the higher orders vanish too in the following way
\begin{equation}
\chi_k(n=1,\gamma=1/2) \rightarrow 0, \; \; \; \;\;\frac{\chi_k(n=1,\gamma=1/2)}{\chi_0(n=1,\gamma=1/2)} \rightarrow {\rm const} \; ,
\end{equation}
with the solution 
$$
\omega_{Odd}=\bar{\alpha}_s \chi(n=1,\gamma=1/2) =0 \; ,
$$
to all orders. Note, that we cannot have the solution $\omega_{Odd}\neq 0$ since this would imply that the ratios
$\frac{\chi_k}{\chi_0}$ should depend on $\bar{\alpha}_s$ which is forbidden by construction.

The situation described above is completely analogous to the properties of the anomalous dimensions. The momentum sum rule for the anomalous dimensions in QCD states that
\begin{eqnarray}
\gamma_{gg}(\omega=1)+2 N_f \gamma_{qg}(\omega=1) & = & 0 \; ,\nonumber \\
\gamma_{gq}(\omega=1)+\gamma_{qq}(\omega=1) & = & 0 \; ,
\label{eq:gammomentum}
\end{eqnarray}
order by order in perturbation theory. We can therefore write the generic  
perturbative expansions for both combinations of anomalous dimensions
\begin{equation}
\Gamma(\omega) = \Gamma_0(\omega) + \sum_{k\ge 1}\alpha_s^k \Gamma_k(\omega)  \;,
\end{equation}
where the momentum sum rule condition gives
$$
\Gamma_0(1)=\Gamma_k(1)=0, 
$$
for all $k$. Here, $\Gamma(\omega)$ denotes one of the combinations, or it can be also the $n_f=0$ part of the $\gamma_{gg}$.
Since the anomalous dimensions are suitable for the description of the collinear limit, which is $\gamma\rightarrow 0$, one can introduce the dual $\gamma$ expansion which reads\footnote{Strictly speaking we are considering here fixed coupling, like in the $N = 4$ SYM case,
but the arguments can be generalized to running coupling.}
\begin{equation}
\label{eq:gammaexp}
\tilde{\Gamma}(\omega,\gamma) = \tilde{\Gamma}_0(\omega) + \sum_{k\ge 1} \gamma^j \tilde{\Gamma}_j(\omega)  \;.
\end{equation}
The corresponding  anomalous dimension can be found from the nonlinear equation\footnote{This procedure was developed  in \cite{Dokshitzer:2005bf,DM}.}
$$
\gamma = \alpha_s \tilde{\Gamma}(\omega,\gamma) \; .
$$
The lowest order condition reads then
$$
\Gamma_0(\omega) =  \tilde{\Gamma}_0(\omega) \; .
$$
One can then relate the coefficients in both expansions to obtain the analog of the (\ref{eq:omegaexp})
\begin{multline}
\tilde{\Gamma}(\omega,\gamma) \, = \, {\Gamma}_0(\omega) +\gamma \frac{\Gamma_1(\omega)}{\Gamma_0(\omega)} \, +\\
+\,\gamma^2 \frac{1}{\Gamma_0(\omega)}\left( \frac{\Gamma_2(\omega)}{\Gamma_0(\omega)}-\left(\frac{\Gamma_1(\omega)}{\Gamma_0(\omega)}\right)^2\right)\,+\,{\cal O}(\gamma^3) \; .
\label{eq:gammaexp1}
\end{multline}
 We see immediately that one encounters exactly the same pattern. When the lowest order vanishes $\Gamma_0(1)=0$, 
 the existence of the $\gamma$ expansion  (\ref{eq:gammaexp},\ref{eq:gammaexp1})    implies  that  $\Gamma(1)=0$.
 We see that the assumption about the  $\gamma$ expansion  is consistent, because we know that all the higher order coefficients have to vanish, due to the momentum sum  rule.

Therefore, strictly following this logic, we are led to the conclusion that,
the validity of the $\omega$ expansion for the kernel eigenvalue in the case of the van-
ishing lowest order imposes strong constraint that the higher orders should
automatically vanish too.

This would imply that the Odderon intercept is equal to unity in each
order of the perturbation theory.

In fact the vanishing of the Odderon solution eigenvalue is most probably
also related to the very same momentum sum rule constraint. As we know, in
the infinite coupling limit the Pomeron trajectory coincides with the graviton
in $AdS_5$ for the case of the $N=4$ SYM. This is manifested by the fact that the intercept becomes 2 in this
limit, or $\omega= 1$ correspondingly. As was demonstrated in \cite{Stasto:2007uv} this is also
related to the momentum sum rule condition. For the Odderon case it seems
though that the conformal spin $n$ and $\omega$ are shifted by $+1$ and $-1$
correspondingly within the same eigenvalue function as the Pomeron. Unlike
the Pomeron case though, the intercept seems to stay equal to 1 for all values
of the coupling.

Considering the arguments above,  we are led to the conjecture that the BFKL equation also satisfies the momentum sum rule in analogy with the anomalous dimensions. The momentum sum rule for the BFKL is however valid for the transverse components of the momenta, whereas the momentum sum rule for the anomalous dimensions is valid for the longitudinal parts of the momenta. 

For the case of the anomalous dimensions we have that
\begin{equation}
\int_0^1 dx\,  x^{\omega=1} (P_{gg}(x)+2\, N_f P_{qg}(x) )=0 \; ,
\end{equation}
which is of course equivalent to the first equation in (\ref{eq:gammomentum}).

For the BFKL equation we should have the analogous condition which can be represented as
\begin{equation}
\chi(n=1,\gamma=1/2) = \int d^2  q  \, \frac{k}{q} \, e^{i\phi} \, K(k,q) \, =\,0 \; .
\end{equation}

The physical interpretation of this statement  is quite clear. In the case of the collinear approximation, the standard integrated parton densities  satisfy the longitudinal momentum sum rule. The sum over  the  longitudinal momenta in the parton densities
is constant, independent of the $Q^2$ evolution. The DGLAP evolution equations  can only redistribute the 
longitudinal momenta but cannot change the longitudinal momentum sum rule constraint. 
In the BFKL case the evolution is in $x$ and the dynamics in the transverse momenta is non-trivial. The transverse momenta can be redistributed in the BFKL evolution  but the overall sum is unchanged which can be expressed in the following condition
\begin{equation}
\frac{\partial}{\partial \ln 1/x} \int \, \frac{d^2 {\bf k}}{k^2} \,{\bf k}\, G(x,{\bf k}) =0 \; ,
\end{equation} 
where $G$ is the $k_T$ unintegrated gluon density.

\section{NLLx with $n=1$ conformal spin}
\label{sec:nllx}
The  hypothesis of  $\omega$ expansion and the BFKL momentum sum rule can be tested by looking into the explicit result for the NLLx eigenvalue at $n=1$ which was calculated in \cite{Fadin:1998py,Kotikov:2000pm} and recently in \cite{Balitsky:2008zz}. We note, that the eigenvalue obtained there is for the case of the Pomeron and the explicit verification for the case of the Odderon
still needs to be performed. The NNLx eigenvalue from \cite{Kotikov:2000pm} for all conformal spins reads
\begin{multline}
\int d^2 q \bigg(\frac{q^2}{k^2}\bigg)^{\gamma-1} e^{in\phi} K(k,q) = \bar{\alpha}_s [\chi_0(n,\gamma)+ \bar{\alpha}_s \chi_1(n,\gamma)] \; , \\
4 \,\chi_1(n,\gamma) = -\frac{b}{2} [\chi_0'(n,\gamma)+\chi^2(n,\gamma)]+\bigg(\frac{67}{9}-\frac{\pi^2}{3}-\frac{10}{9}
\frac{n_f}{N_c^3}\bigg) \chi_0(n,\gamma)\\
+6 \zeta(3) -\chi''_0(n,\gamma)+F(n,\gamma)-2\Phi(n,\gamma)-2\Phi(n,1-\gamma) \; ,
\label{eq:flnll}
\end{multline}
where 
\begin{multline}
\Phi(n,\gamma) = \int_0^1 \frac{dt}{1+t} t^{\gamma-1+\frac{n}{2}} \left\{ \frac{\pi^2}{12}-\frac{1}{2} \psi^{\prime}(\frac{1+n}{2})-{\rm Li}_2(t)-{\rm Li}_2(-t)\right. \\ \left.
-\ln t\bigg(\psi(1+n)-\psi(1)+\ln(1+t)+\sum_{k=1}^{\infty} \frac{(-t)^k}{k+n} \bigg) -\sum_{k=1}^{\infty} \frac{t^k}{(k+n)^2}[1-(-1)^k] \right\} \; ,
\label{eq:phi}
\end{multline}
and the non-analytic part of the kernel is defined
$$
F(n,\gamma) = \delta_{0n} F_1 + \delta_{2n} F_2 \, .
$$
with functions $F_1,F_2$ defined in \cite{Kotikov:2000pm}. The non-analytic part is obviously zero for $n=1$.
For $n=1$ the sums in (\ref{eq:phi}) can be performed explicitly and yield
\begin{equation}
\sum_{k=1}^{\infty} \frac{(-t)^k}{k+1} =\frac{1}{t} (-t + \ln(1+t)) \; , 
\end{equation}
and
\begin{equation}
\sum_{k=1}^{\infty} \frac{t^k}{(k+1)^2}[1-(-1)^k]=\frac{1}{t}({\rm Li}_2(t)+{\rm Li}_2(-t) ) \; .
\end{equation}
Using above expressions one can recast (\ref{eq:phi}) for $n=1$ into the form
\begin{multline}
\Phi(1,\gamma) \; = \; -\frac{1}{6 (-1+2 \gamma)^3}\left[48- 24 \psi(1) (1-2 \gamma)+
\pi^2 ( 1-2 \gamma  )^2+\right.\\ \left.
24 (1-2 \gamma) \psi(\frac{1}{2}+\gamma)-3 (1-2 \gamma)^2 \psi_1(-\frac{1}{4}+\frac{\gamma}{2})+3 (1-2 \gamma)^2 \psi_1(\frac{1}{4}+\frac{\gamma}{2})\right] \;.
\label{eq:phi1}
\end{multline}
Investigating expression (\ref{eq:flnll}) we immediately see that the terms proportional to $b$ and $\chi_0$ vanish at $n=1,\gamma=\frac{1}{2}$, whereas from (\ref{eq:phi1}) we have that $\Phi(1,\gamma=\frac{1}{2})=\frac{\zeta(3)}{2}$. The second derivative is $\chi_0''(1,1/2)=4\zeta(3)$ and the whole next-to-leading correction vanishes
$$
\chi_1(n=1,\gamma=1/2) =0 \;, 
$$
which is consistent with the $\omega$ expansion hypothesis. We note that, the result obtained in \cite{Balitsky:2008zz} does not vanish as it differs from the one cited above by constant term proportional to $2\zeta(3)$ in $4\chi_1$.

\begin{figure}[t]
\centerline{\includegraphics[width=0.45\textwidth]{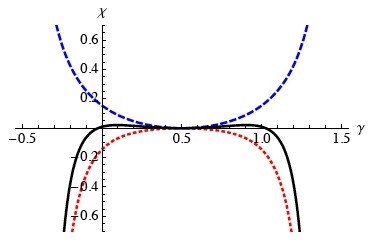}\hfill \includegraphics[width=0.45\textwidth]{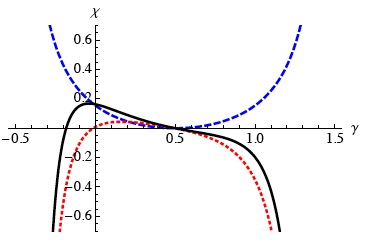}}
\caption{Left: the LLx eigenvalue with $n=1$ (dashed blue curve) as compared to the NLLx eigenvalue  (dotted red). The sum of LLx and NLLx is solid black curve. $\alpha_s N_c/\pi=0.2$ and $b=0$. Right: the same for running coupling case. }
\label{eq:nllplot}
\end{figure}

We illustrate the $\chi_1(n=1,\gamma)$ together with $\chi_0(n=1,\gamma)$ in left plot in  Fig.~\ref{eq:nllplot}, for $b=0$ and $\bar{\alpha}_s =0.2$.  On the right plot we show the case with the running coupling. The eigenvalue is asymmetric due to the terms proportional to $\beta$. A characteristic feature is the fact that the diffusion becomes very small when NLL corrections are included.

 
\section{Odderon with kinematical constraints}
\label{sec:oddkc}
As an example of the application of the $\omega$ expansion we will investigate what happens to the kernel eigenvalue and the intercept in the case when we include the shifts of the collinear poles in the kernel eigenvalue for $n=1$ case.
The shifts originate from the kinematical constraint \cite{Kwiecinski:1996td,Andersson:1995ju} and are
an important ingredient of the resummation procedure at small x.
 We will call this case the Odderon with the kinematical constraints.
Let us take the following  kernel
\begin{equation}
\tilde{\chi}(n=1,\gamma,\omega) \, = \, 2 \psi(1) - \psi(\gamma + \frac{1}{2} + \frac{\omega}{2})-\psi(1-\gamma+\frac{1}{2}+\frac{\omega}{2}) \;,
\label{eq:oddkc}
\end{equation}
which for $\omega=0$ reduces to the leading order BLV Odderon kernel 
$$
\chi_0(n=1,\gamma)=\tilde{\chi}(n=1,\gamma,\omega=0) \; .
$$
It is clear that the shifts do not change the value of the intercept in this case
\begin{equation}
\omega_0 = \bar{\alpha}_s \chi_0(n=1,\gamma=\frac{1}{2}) = \bar{\alpha}_s \tilde{\chi}(n=1,\gamma=\frac{1}{2},\omega=0)  \, = \, 0 \; . 
\end{equation}
The expansion in powers of $\omega$ proceeds as follows
\begin{multline}
\omega =  \bar{\alpha}_s \tilde{\chi}(1,\gamma,\omega) =\\
=\bar{\alpha}_s \tilde{\chi}_0^{(0)}(1,\gamma,0) + \bar{\alpha}_s \omega \tilde{\chi}_0^{(1)}(1,\gamma,0)+\bar{\alpha}_s \omega^2 \tilde{\chi}_0^{(2)}(1,\gamma,0) + \dots \; ,
\end{multline}
where
$$
\tilde{\chi}^{(k)}(1,\gamma,0) = \frac{1}{k!}\frac{\partial^k \tilde{\chi}(1,\gamma,\omega)}{\partial \omega^k}|_{\omega=0} \; .
$$

The second order in $\alpha_s$ reads then
\begin{equation}
\omega_1(\gamma) = \bar{\alpha}_s \tilde{\chi}_0^{(0)}+\bar{\alpha}_s^2 \tilde{\chi}_0^{(0)}\tilde{\chi}_0^{(1)} \; ,
\label{eq:first}
\end{equation}
where we dropped the explicit arguments.
The third order in $\bar{\alpha}_s$ can be obtained via next iteration and is equal
\begin{equation}
\omega_2(\gamma) = \bar{\alpha}_s \tilde{\chi}_0^{(0)}+\bar{\alpha}_s^2 \tilde{\chi}_0^{(0)}\tilde{\chi}_0^{(1)}\,
 + \;  \bar{\alpha}_s^3 (\tilde{\chi}_0^{(0)})^2  \tilde{\chi}_0^{(2)} \, + \, \bar{\alpha}_s^3 \tilde{\chi}_0^{(0)}  (\tilde{\chi}_0^{(1)})^2 \; .
\label{eq:second}
\end{equation}
Clearly at each order of $\alpha_s$ the solution to the intercept is again zero, since in (\ref{eq:first}) and (\ref{eq:second}) all the coefficients are proportional to $\tilde{\chi}_0^{(0)}$ which vanishes when $\gamma=1/2$.
On the other hand the diffusion coefficient receives non-trivial correction due to the shifts. Using (\ref{eq:first}) 
and (\ref{eq:second}) we can write down the expansion for the diffusion 
$$
D=\bar{\alpha}_s \, D_0+\bar{\alpha}_s^2 \, D_1 + \bar{\alpha}_s^3 \, D_2 \; + \; \dots \;,
$$
with the first three coefficients equal
$$
D_0  = -\psi_2(1), \; \; D_1= \psi_2(1) \frac{\pi^2}{6}, \; \; D_2 = -\psi_2(1) \frac{\pi^4}{36} \; .
$$
where $\psi_2(1)=-2\zeta(3)$.

\begin{figure}[t]
\centerline{\includegraphics[width=0.45\textwidth]{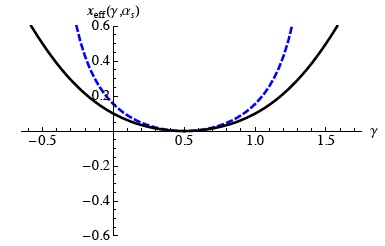}\hfill \includegraphics[width=0.45\textwidth]{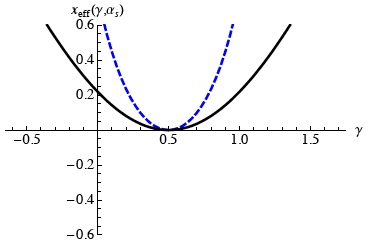}}
\caption{Left: the LLx eigenvalue with $n=1$ (dashed blue curve) as compared to the resummed eigenvalue  obtained from solution to eigenvalue equation with Eq.\ref{eq:oddkc} (solid black curve) for $\alpha_s N_c/\pi =0.2$. Right: the same for  $\alpha_s N_c/\pi =1.0$ }
\label{fig:eigenvalueodd}
\end{figure}

We have solved the implicit  equation (\ref{eq:implicit}) for $\omega$ with the modified kernel eigenvalue (\ref{eq:oddkc}) numerically.
The effective eigenvalue as a function of $\gamma$ is shown in Fig.~\ref{fig:eigenvalueodd}
together with the leading order result. We have chosen two values of the coupling: small $\alpha_s N_c/\pi=0.2$ (left panel) and large $\alpha_s N_c/\pi=1.0$ (right panel). As expected the zero value at $\gamma=1/2$ is the same, while the shape of the eigenvalue is affected significantly.

 
The resummed second derivative is again much smaller than in the LLx
case. We have checked that the second derivative in the shifted case saturates
at the limit of about $2.8$ for very large values of the coupling constant.

\section*{Summary}
\label{sec:summary}

We have constructed a general argument, based on the validity of the $\omega$ expansion that
if the leading order eigenvalue vanishes, then it implies that the higher order terms automatically vanish too.
This is strongly supported by the analogous pattern observed for the anomalous dimensions which in the latter case is just a consequence of the momentum sum  rule.

This fact strongly suggests that the BLV Odderon intercept  is unchanged in the higher orders of perturbation theory. However, for this to be established an explicit evaluation of the Odderon in the NLLx is necessary.

The $\chi_1(1,\gamma)$ coefficient vanishes in the case of the calculations performed in momentum space,
which is consistent with the $\omega$ expansion. However, it does not vanish in the dipole approach, which indicates that some modifications to the dipole kernel might be necessary in order to ensure the compatibility with the $\omega$ expansion.
Based on the analogy  with the anomalous dimensions we conjectured that the BFKL equation should also satisfy the momentum sum rule for the transverse momenta.

We have also analyzed the modification of the lowest order Odderon kernel eigenvalue due to the shifts of poles, which are originate  from the kinematical effects. We found that the eigenvalue is still zero and that the diffusion
is significantly reduced with respect to the lowest order case.


\section*{Acknowledgments}
I thank Stanley Brodsky for the question about the resummation in the
Odderon case.
This work
was supported by the  Polish Ministry of Education grant No.\ N202 249235 and 
by the Alfred P. Sloan Foundation.

\end{document}